\documentclass[reprint,showpacs,aps,prl,superscriptaddress,amsmath,amsfonts,amssymb]{revtex4-1}

\usepackage[]{graphicx}
\usepackage{amsmath}
\usepackage{times}
\usepackage{bm}
\usepackage{units}
\usepackage{ulem}
\usepackage{color}
\renewcommand{\bm }{\mathbf}

\newcommand{\comment}[1]{}

\begin{document}

\title{Impact of Auger processes on carrier dynamics in graphene }

\author{Torben Winzer}
\email[]{t.winzer@mailbox.tu-berlin.de}
\author{Ermin Mali\'{c}}

\affiliation{Institut f\"ur Theoretische Physik, Technische Universit\"at Berlin,
  Hardenbergstr. 36, 10623 Berlin, Germany}

\begin{abstract}
The linear, gapless bandstructure of graphene provides ideal conditions for Auger processes. They are of great importance for fundamental research and technological applications, since they qualitatively change the carrier dynamics. Time-, momentum-, and angle-resolved microscopic calculations reveal an efficient impact excitation giving rise to a significant multiplication of optically excited carriers and a remarkable Coulomb-induced carrier cooling effect. We present an analytic expression for the carrier multiplication predicting ideal conditions for its still missing experimental observation.
\end{abstract}

\maketitle


Graphene consists of a single layer of carbon atoms arranged in a honeycomb crystal lattice.
Its reduced dimensionality and the unusual linear, gapless bandstructure leads to unique electronic and optical properties suggesting its application in various opto-electronic devices \cite{geim07,ferrari10b,neto09}. The key prerequisite for the realization of such applications is a better microscopic understanding of the temporal evolution of the carrier density in a non-equilibrium \cite{ferrari10b}. In this context, it is of great importance to study the efficiency of Auger scattering channels, which change the carrier density by bridging the valence and the conduction band. In contrast to  semiconductors
having a bandgap and a parabolic bandstructure, graphene shows efficient Auger-type scattering \cite{auger_rana07,auger_winzer10,auger_avouris11}.

In the last years, a number of theoretical \cite{butscher07,auger_rana07, auger_winzer10,auger_avouris11,relax_PRB} and experimental studies \cite{relax_MBI,relax_PRL,relax_Norris,wang10_dts,auger_yuri11,HEINZ} were performed with the aim to obtain a detailed understanding of the carrier relaxation dynamics in graphene. However, only a few of these studies consider the role of Auger scattering channels \cite{auger_rana07, auger_winzer10,auger_yuri11}. 
In this work, we obtain new insights into the influence of these channels on the carrier density, the carrier temperature, and the chemical potential in optically excited graphene.
Based on a microscopic approach offering access to time-, momentum-, and angle-resolved carrier relaxation dynamics, we predict (i) a significant carrier multiplication (CM) even in the presence of directly competing phonon-induced recombination,  and (ii) a counter-intuitive Coulomb-induced carrier cooling effect, a process widely overseen in current literature. 
We further address the efficiency of Auger-type channels by deriving an analytic expression for the strength of the CM reflecting its dependence on the pump fluence, temperature, and excitation energy. This allows us to give the optimal conditions for its still missing experimental observation in graphene. 

 \begin{figure}[t!]
  \begin{center}
\includegraphics[width=0.95\linewidth]{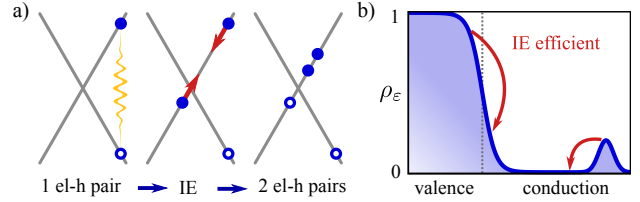}
  \end{center}
  \caption{(a) Illustration of the impact excitation (IE) increasing the number of charge carriers. Two electron-hole pairs are generated by the absorption of a single photon - a process called carrier multiplication (CM). (b) Schematic carrier distribution showing the efficiency of IE at the band edge and displaying the possible scattering partners provided by the optically excited non-equilibrium distribution. } \label{fig1} 
\end{figure}

Carrier multiplication is defined as the ratio between the total number of generated and the initially optically excited carrier-hole pairs, reflecting the enhancement of the quantum yield due to many-particle effects.
Under certain conditions, the absorption of a single photon can generate multiple electron-hole pairs, cp. Fig. \ref{fig1}.  It has already been observed experimentally and studied theoretically in quantum dots \cite{Auger_QD,Franz} and carbon nanotubes \cite{auger_cnt_exp,auger_cnt_dts,auger_cnt_theo,auger_cnt_transit}. In graphene however, an efficient CM has been predicted theoretically \cite{auger_winzer10}, but has not yet been directly measured in experiments. Recent studies on CVD-grown multilayer graphene already indicate that Coulomb-scattering leads to a change in the carrier density \cite{auger_yuri11}.

The process of CM is driven by Auger scattering consisting of two inverse processes: impact excitation (IE) and Auger recombination (AR). Figure \ref{fig1}a illustrates IE, where an optically excited electron scatters down to an energetically lower state. At the same time, another electron is excited from the valence into the conduction band. As a result, we obtain two electron-hole pairs by the absorption of a single photon.  In contrast, the inverse process of Auger recombination gives rise to a decrease in the charge carrier density (reverse arrows in Fig. \ref{fig1}b).

Our time-resolved calculations are based on the graphene Bloch-equations \cite{relax_PRB}, allowing for a consistent consideration of carrier-carrier and carrier-phonon scattering on a microscopic footing  \cite{haug, rossi02}. The relaxation dynamics is described by a coupled set of differential equations for (i) the occupation probability $\rho^{\lambda}_{\bf k}$ in the state $\bf k$ for electrons and holes ($\lambda=c,h$), (ii) the microscopic polarization $p_{\bf k}$, which is a measure for the transition probability between the bands,  and (iii) the phonon occupation $n^j_{\bf q}$ for different optical and acoustic modes $j$:
\begin{equation}\label{GBE}
 \begin{split}
  \dot{\rho}_{\bf k}^{\lambda}&=2\Im\left[\Omega_{\bf k}^{vc*}p_{\bf k}\right]+\dot{\rho}_{\bf k}^{\lambda}\Big|_{scat},\\
\dot{p}_{\bf k}&=\left[i\Delta\omega_{\bf k}-i\Omega_{\bf k}^{\lambda\lambda}-\gamma_{\bf k}\right]p_{\bf k}-i\Omega_{\bf k}^{vc}\left[\rho_{\bf k}^c-\rho_{\bf k}^v\right]+\mathcal{U}_{\bf k},\\
\dot{n}_{\bf q}^j&=-\gamma_{ph}\left[n_{\bf q}^j-n_B\right]+\Gamma_{j,{\bf q}}^{em}\left[1+n_{\bf q}^j\right]-\Gamma_{j,{\bf q}}^{abs}n_{\bf q}^j.
 \end{split}
\end{equation}
Besides the transition frequency $\Delta\omega_{\bf k}$, the phenomenological phonon decay rate $\gamma_{ph}$ \cite{phononlifetime}, and the Bose-Einstein distribution $n_B$ as the initial condition for the phonon occupation, all terms in in Eq. (\ref{GBE}) are explicitly time-dependent. The Rabi frequency $\Omega_{\bf k}^{vc}$ describes the coupling to the exciting vector potential. $\Omega_{\bf k}^{\lambda\lambda}$ is the corresponding intraband contribution. The many-particle interactions are treated within the second-order Born-Markov approximation resulting in emission $\Gamma_{j,{\bf q}}^{em}$ and absorption rates $\Gamma_{j,{\bf q}}^{abs}$ for the phonons. For the microscopic polarization $p_{\bf k}(t)$ we obtain diagonal $\gamma_{\bf k}(t)$ and off-diagonal dephasing $\mathcal{U}_{\bf k}(t)$ and for the carrier occupation probability $\rho_{\bf k}^{\lambda}$ a Boltzmann-like scattering equation \cite{relax_PRB}:
\begin{equation}
 \dot{\rho}^{\lambda}_{\bf k}\Big|_{scat}\!=\!\sum\limits_{\lambda_1,{\bf k}'}\!\mathcal{W}_{{\bf k}'\rightarrow{\bf k}}^{\lambda_1\rightarrow\lambda}\rho^{\lambda_1}_{{\bf k}'}(1-\rho^{\lambda}_{\bf k})-\mathcal{W}_{{\bf k}'\leftarrow{\bf k}}^{\lambda_1\leftarrow\lambda}(1-\rho^{\lambda_1}_{{\bf k}'})\rho^{\lambda}_{\bf k}.\label{eq_scatt}
\end{equation}
In the case of the Coulomb interaction, the probability to scatter from the state$(\lambda_1,{\bf k}')$ to $(\lambda,{\bf k})$ is given by:
\begin{equation}
\mathcal{W}_{{\bf k}'\rightarrow{\bf k}}^{\lambda_1\rightarrow\lambda}=\sum\limits_{\lambda_2,\lambda_3,{\bf q}}\mathcal{V}^{\lambda,\lambda_1,\lambda_2,\lambda_3}_{{\bf k},{\bf k}',{\bf q}}(1-\rho^{\lambda_3}_{{\bf k}'+{\bf q }})\rho^{\lambda_2}_{{\bf k}+{\bf q }},\label{eq_rate}
\end{equation}
where $\mathcal{V}^{\lambda,\lambda_1,\lambda_2,\lambda_3}_{{\bf k},{\bf k}',{\bf q}}$ contains the screened Coulomb matrix element as well as a delta-function ensuring energy conservation \cite{relax_PRB}. The simultaneous conservation of momentum and energy restricts Auger-type processes (IE, AR) to parallel scattering along the Dirac cone \cite{auger_rana07}, cp. Fig. \ref{fig1}a.  The efficiency of impact excitation (IE) $\mathcal{V}^{cvcc}_{{\bf k},{\bf k}',{\bf q}}$ can be understood in terms of two sub-processes denoted in Eqs. (\ref{eq_scatt}) and (\ref{eq_rate}). The primary sub-process (Eq. \ref{eq_scatt}) ${\bf k}'\rightarrow{\bf k}$ bridges both bands and prefers an abrupt gradient in the occupation, which is given by a thermal distribution around the Dirac-point , cp. Fig. \ref{fig1}b. Here, the inverse process of Auger recombination (AR) is strongly suppressed by Pauli-blocking. The sum in Eq. \ref{eq_rate} includes all secondary sub-processes ${\bf k}+{\bf q}\rightarrow{\bf k}'+{\bf q}$ matching in energy and momentum. Note that a non-equilibrium carrier distribution is needed to initiate Auger-type scattering, cp. also Fig. \ref{fig1}b. As a result, the efficiency of IE and AR is given by the corresponding Coulomb matrix elements and the occupation probability of all involved states.

\begin{figure}[t!]
  \begin{center}
\includegraphics[width=0.95\linewidth]{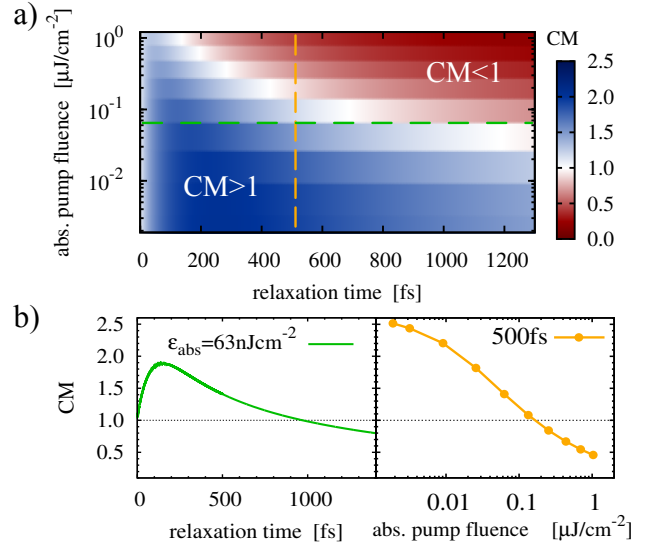}
  \end{center}
  \caption{(a) Temporal evolution of the Auger-induced carrier multiplication (CM) for different pump fluences at $300$~K. The applied optical pulse has an photon energy of $1.5$~eV and a duration of 10 fs corresponding to recent experiment \cite{relax_MBI}. CM is efficient and stable on a picosecond timescale at low pump fluences. (b) Time-dependent CM at a fixed pump fluence and (c) fluence-dependent of CM at a fixed time corresponding to the two straight lines in the contour plot.  } \label{fig2} 
\end{figure}

The numerical evaluation of the Bloch equations enables us to track the excited carriers on their way to equilibrium. Our calculations reveal that the relaxation dynamics is characterized by a significant carrier multiplication, as shown in Fig. \ref{fig2}. 
Since the efficiency of Coulomb-induced scattering processes strongly depends on the number of available charge carriers in the system, we study the temporal evolution of CM as a function of the optically-induced pump fluence $\varepsilon_{opt}$, cp. Fig. \ref{fig2}a.  
The larger the pump fluence, the more scattering partners are available and the faster is the relaxation dynamics. As a result, a thermalized distribution is reached already within the first tens of fs and the impact of Auger processes is restricted to a very short time frame. The direct consequence is a significantly reduced CM for large pump fluences. Moreover, the carrier-phonon scattering channels reduce CM for two reasons: (i) the carrier density  decreases via phonon-induced recombination processes and (ii) there is a direct competition between Auger and phonon processes, where an excited electron can either perform Auger scattering or reduce its energy by emitting a phonon. Figure \ref{fig2} maps the temporally resolved interplay between Coulomb- and phonon-induced channels at different pump fluences. Two different regions characterized by $\rm{CM}>1$ (blue area) and $\rm{CM}<1$ (red area) can be clearly seen. For a better quantification, two exemplary lines of the contour plot are depicted in Fig. \ref{fig2}b showing the cross-over between both regions. The carrier multiplication as a function of time at a fixed pump fluence illustrates the efficiency of phonon-induced recombination. On the other side, the fluence-dependent cross-over reflects the accelerated Coulomb scattering at higher carrier densities. 
We predict a CM of up to a value of 2.5 at low pump fluences. Here, $\rm{CM}>1$ remains stable on a ps time scale confirming the potential of graphene for applications in opto-electronics.

Carrier multiplication can be treated within the quasi-equilibrium quantum statistics in consideration of a purely Coulomb-induced relaxation dynamics. Energy dissipation and recombination due to carrier-phonon scattering reduces CM, however, it does not change its qualitative behavior. The carrier density $n$ and the energy density $\varepsilon$ of the electronic system are determined by the microscopic carrier occupation $\rho^\lambda_{\bm k}$ via
\begin{equation}
n=\frac{\sigma_s\sigma_v}{A}\sum\limits_{\lambda{\bf k}}\rho_{\bf k}^{\lambda}\;\;\;\mbox{and}\;\;\;\varepsilon=\frac{\sigma_s\sigma_v}{A}\sum\limits_{\lambda{\bf k}}\varepsilon_{\bf k}^{\lambda}\rho_{\bf k}^{\lambda},\label{density}
\end{equation}
where $\sigma_s$ and $\sigma_v$ stand for the spin and valley degeneracy, respectively. Furthermore, $A$ is the graphene area and $\varepsilon_{\bf k}^{\lambda}$ denotes the electronic bandstructure. In equilibrium $\rho_{\bf k}^{\lambda}$ is given by a Fermi distribution and both carrier and energy density are functions of the temperature $T$ and the chemical potential  $\mu$, and vice versa. Therefore, it is possible to completely determine the Coulomb-induced final state of the system after an optical excitation. In the case of a conventional semiconductor, where Auger-processes are negligible, the absorbed pump fluence $\varepsilon_{opt}$ and the optically injected carrier density $n_{opt}=2\varepsilon_{opt}/\hbar\omega$ are both conserved and form new separate Fermi distributions in the conduction and the valence band. In contrast, in graphene only the energy is conserved, whereas the carrier density is a variable quantity giving rise to a single Fermi distribution in both bands. In particular, in the case of undoped graphene, the symmetry between electrons and holes results in a vanishing chemical potential for the equilibrated system. Consequently, in this situation the carrier and energy density are determined only by the electronic temperature. Assuming a linear energy dispersion $\varepsilon_{\bf k}^{\lambda}=\pm\nu_F |\bf k|$, Eq. (\ref{density}) yields $n(T)=c_1T^2$ and $\varepsilon(T)=c_2T^3$ with the coefficients $c_1=\pi k_B^2/3\nu_F^2$ and $c_2=6\zeta(3)k_B^3/\pi \nu_F^2,$ where $\zeta$ denotes the Zeta-function \cite{butscher07} . Starting with an initial thermal distribution at the temperature $T_0$ and assuming again an optical excitation, we can successively calculate the final state of the system $f$ in terms of the energy density $\varepsilon_f=\varepsilon(T_0)+\varepsilon_{opt}$, the temperature $T_f(\varepsilon_f)$, and the carrier density $n_f(T_f)$. Finally, we obtain an analytic expression, describing the carrier multiplication as a function of the initial temperature $T_0$, the absorbed pump fluency $\varepsilon_{opt}$, and the excitation energy $\hbar\omega$:
\begin{equation}
\label{CM_formula}
{\rm{CM}}=\frac{n_f-n(T_0)}{n_{opt}-n(T_0)}=\frac{c_1\hbar\omega}{2\varepsilon_{opt}}\bigg[\big(T_0^3+\frac{\varepsilon_{opt}}{c_2}\big)^{2/3}-T_0^2\bigg].
\end{equation}
Note that this relation describes the final state of the carrier system due to the carrier-carrier scattering and is independent of the relaxation time. In particular, Eq. (\ref{CM_formula}) does not include the phonon-induced reduction of CM, which cannot be statistically treated. Nevertheless, the presented analytic expression outlines a systematic estimation of the efficient regime for a high carrier multiplication.

\begin{figure}[t!]
  \begin{center}
\includegraphics[width=\linewidth]{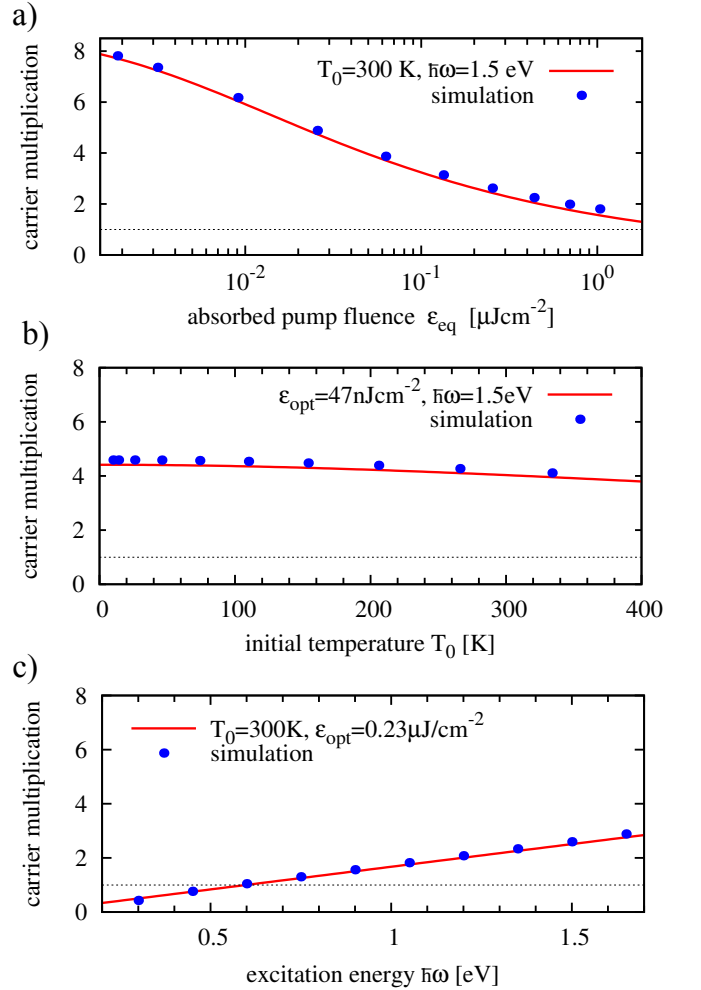}
  \end{center}
  \caption{Analytically obtained purely Coulomb-induced carrier multiplication (red solid line) is compared to the numerical solution from Bloch equations (blue dots) for different (a) absorbed pump fluences $\varepsilon_{opt}$, (b) initial temperatures $T_0$, and (c) excitation energies $\hbar \omega$.} \label{fig3} 
\end{figure}

The different dependences of the carrier multiplication ${\rm{CM}}(\varepsilon_{opt},T_0,\hbar\omega)$ are illustrated in Figure \ref{fig3}. The comparison between analytic and numerical results shows an excellent agreement. The highest CM is found for small pump fluences, small initial temperatures, and high excitation energies. The influence of the pump fluence has already been discussed in the previous section - the smaller the pump fluence the larger is CM, as shown in Fig. \ref{fig3}a. Now, this dependence is quantified by Eq. \ref{CM_formula}, in particular for small initial temperatures, CM is proportional to $\varepsilon_{opt}^{-1/3}$. For a fixed excitation, CM shows only weak dependence of the initial temperature $T_0$, as displayed in Fig. \ref{fig3}b. At higher temperatures, it slightly decreases resulting from a reduced asymmetry between the Auger-type processes of IE and AR. The higher the temperature, the smoother is the Fermi function and the smaller is the suppression of AR due to the Pauli blocking. However, compared with the pump fluence, the influence of the initial temperature is small in the range between $\unit[10]{K}$ and $\unit[400]{K}$. Finally, the appearing CM scales linearly with the excitation energy $\hbar \omega$, as predicted by the analytic expression and shown in Fig. \ref{fig3}c. This can be understood intuitively, since the further away from the Dirac point a carrier is excited, the more IE it can induce on its way back to equilibrium. If the non-equilibrium is located close to the Dirac point, i.e. in the case of small excitation energies, the scattering probabilities differ qualitatively. In contrast to the scattering event depicted in Fig. \ref{fig1}b, both sub-processes can directly be initiated by the optically excited carries. This leads to an enhanced efficiency for AR, which can even result in $\rm{CM}<1$, cp. Fig. \ref{fig3}c.

In conclusion, ideal condition for an experimental observation of CM are a relatively small pump fluence, a low initial temperature, and a high excitation energy at the same time. The comparison with Fig. \ref{fig2} shows that the analytically predicted efficient regime for CM is also predominant in the presence of carrier-phonon scattering. Thus, the presented analytic expression gives crucial insights into the dependences of CM and helps to estimate its most efficient regime, also under realistic conditions. Note that deviations from the linear bandstructure may reduce the CM. However, we expect non-Markov processes to compensate this effect, since they soften the strict energy conservation \cite{haug}.

\begin{figure}[t!]
  \begin{center}
\includegraphics[width=0.9\linewidth]{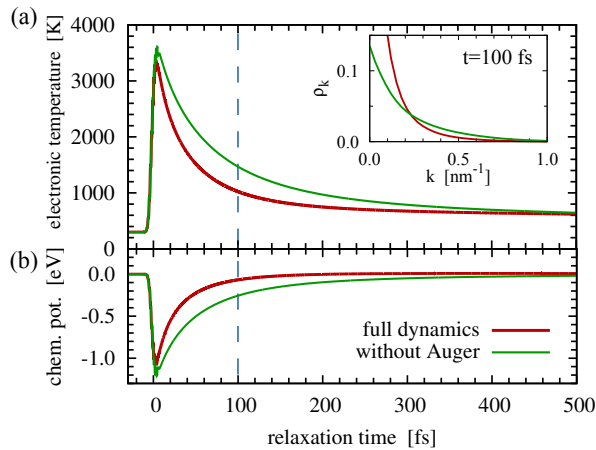}
  \end{center}
  \caption{Temporal evolution of the (a) temperature  and (b) chemical potential of an optically excited carrier system illustrating the contribution of Auger-induced processes to carrier cooling and faster equilibration of the chemical potential, respectively.  The inset shows the momentum-resolved carrier distribution after $\unit[100]{fs}$ with and without Auger channels.} \label{fig4} 
\end{figure}

Another interesting and counterintuitive aspect of the Auger processes is a Coulomb-induced carrier cooling effect. In the case of pure carrier-carrier scattering, where the total energy of the carrier system is conserved, the electronic temperature is a monotonous function of the carrier density. Adding carriers to the system via IE, results in a decreased temperature. In contrast, AR reduces the number of carriers accounting for a carrier heating. We determine the electronic temperature as a function of the carrier and energy density, which are obtained by the numerical evaluation of the Bloch equations including carrier-carrier as well as carrier-phonon scattering channels \footnote{the temperature and the chemical potential are only well-defined for equilibrated carrier distributions. Thus, during the pulse and in the first 50 fs of the relaxation dynamics, they can only be interpreted as statistic quantities reflecting the ratio of the carrier and the energy density.}. Figure \ref{fig4}a shows the temporal evolution of the temperature, illustrating that the predominant IE indeed contributes to a more efficient carrier cooling: Switching off the Auger channels slows down the decrease in the carrier temperature, resulting in a difference of about $\unit[400]{K}$ after $\unit[100]{fs}$. The enhanced cooling is also accompanied by a faster equilibration of the chemical potential, since the same energy is shared by an increasing number of available carriers, cp. Fig.\ref{fig4}b. This is further shown by the corresponding carrier distributions in the inset. The presence of Auger scattering leads to higher occupations at small energies, whereas the neglect of IE and AR results in a broader carrier distribution.

In conclusion, we have microscopically investigated the role of Auger processes on the ultrafast carrier relaxation dynamics in graphene. We predict a significant carrier multiplication and an efficient carrier cooling due to the strong impact excitation. In spite of the directly competing phonon-induced processes, we find a multiplication of carriers by a factor of 2.5 confirming the potential of graphene as a new material for high-efficiency photo-devices. The presented analytic expression gives new insights into Auger scattering and guides new experiments on graphene and related structures.

We acknowledge the financial support from the Deutsche Forschungsgemeinschaft
through SPP 1459. E. M. thanks the Einstein Foundation Berlin.

\end{document}